\documentclass[final,onefignum,onetabnum]{siamart171218}

\usepackage{braket,amsfonts}

\usepackage{array}
\newcommand{\cmmnt}[1]{}
\usepackage[caption=false]{subfig}
\usepackage{pgfplots}
\newsiamthm{claim}{Claim}
\newsiamremark{remark}{Remark}
\newsiamremark{hypothesis}{Hypothesis}
\crefname{hypothesis}{Hypothesis}{Hypotheses}

\usepackage{graphicx,epstopdf}

\Crefname{ALC@unique}{Line}{Lines}

\usepackage{amsopn}

\usepackage{xspace}
\usepackage{bold-extra}
\usepackage[most]{tcolorbox}

\colorlet{texcscolor}{blue!50!black}
\colorlet{texemcolor}{red!70!black}
\colorlet{texpreamble}{red!70!black}
\colorlet{codebackground}{black!25!white!25}

\lstdefinestyle{siamlatex}{%
  style=tcblatex,
  texcsstyle=*\color{texcscolor},
  texcsstyle=[2]\color{texemcolor},
  keywordstyle=[2]\color{texemcolor},
  moretexcs={cref,Cref,maketitle,mathcal,text,headers,email,url},
}

\tcbset{%
  colframe=black!75!white!75,
  coltitle=white,
  colback=codebackground, % bottom/left side
  colbacklower=white, % top/right side
  fonttitle=\bfseries,
  arc=0pt,outer arc=0pt,
  top=1pt,bottom=1pt,left=1mm,right=1mm,middle=1mm,boxsep=1mm,
  leftrule=0.3mm,rightrule=0.3mm,toprule=0.3mm,bottomrule=0.3mm,
  listing options={style=siamlatex}
}

\newtcblisting[use counter=example]{example}[2][]{%
  title={Example~\thetcbcounter: #2},#1}

\newtcbinputlisting[use counter=example]{\examplefile}[3][]{%
  title={Example~\thetcbcounter: #2},listing file={#3},#1}

\patchcmd\newpage{\vfil}{}{}{}
\begin{tcbverbatimwrite}{tmp_\jobname_header.tex}
\title{The Dynamics of Bilateral Olfactory Search and Navigation\thanks{Submitted to the editors 06/03/2019
\funding{BE and NR were funded by National Science Foundation Grant PHY1555916. JDV and SDB were funded by National Science Foundation Grant IOS 1555891 }}}

\author{Nour Riman\footnotemark[2]\ \footnotemark[5] 
\and Jonathan D. Victor\footnotemark[3] 
\and Sebastian D. Boie\footnotemark[3] 
\and Bard Ermentrout\footnotemark[4]\ \footnotemark[5] }

\headers{Analysis of bilateral search}{Nour Riman, et. al}
\end{tcbverbatimwrite}
\input{tmp_\jobname_header.tex}

\ifpdf
\hypersetup{ pdftitle={The Dynamics of Bilateral Olfactory Search and Navigation } }
\fi

\begin{document}
\maketitle

\renewcommand{\thefootnote}{\fnsymbol{footnote}}

\footnotetext[2]{Neuroscience Institute, Carnegie Mellon University, Pittsburgh, Pennsylvania 15260 (\email{nourr@andrew.cmu.edu}).} 
\footnotetext[3]{Feil Family Brain and Mind Research Institute, Weill Cornell Medical College, New York, New York 10065 (\email{jdvicto@med.cornell.edu} , \email{sebboie@mailbox.org}).}
\footnotetext[4] {Department of Math, University of Pittsburgh, Pittsburgh, Pennsylvania 15260 (\email{bard@pitt.edu}).}
\footnotetext[5] {Center for the Neural Basis of Cognition, Carnegie Mellon University, Pittsburgh, Pennsylvania 15260.}

\captionsetup[figure]{labelfont={bf},labelformat={default},labelsep=period,skip=0pt}
\captionsetup[subfigure]{labelformat=simple, labelsep=period,singlelinecheck=off,justification=raggedright,skip=0pt}

% REQUIRED
\begin{abstract}
Animals use stereo sampling of odor concentration to localize sources and follow odor trails. We analyze the dynamics of a bilateral model that depends on the simultaneous comparison between odor concentrations detected by left and right sensors. The general model consists of three differential equations for the positions in the plane and the heading.  When the odor landscape is an infinite trail, then we reduce the dynamics to a planar system whose dynamics have just two fixed points. Using an integrable approximation (for short sensors) we estimate the basin of attraction. In the case of a radially symmetric landscape, we again can reduce the dynamics to a planar system, but the behavior is considerably richer with multi-stability, isolas, and limit cycles.  As in the linear trail case, there is also an underlying integrable system when the sensors are short.  In odor landscapes that consist of multiple spots and trail segments, we find periodic and chaotic dynamics and characterize the behavior on trails with gaps and that turn corners.
\end{abstract}

% REQUIRED

% REQUIRED

\section{Introduction}
\label{sec:intro}
Animals use olfactory cues to navigate through their environment in order to find food, encounter mates, avoid predators and locate their home. This requires an ability to both localize odor sources and follow odor trails. To localize odor, animals have been observed to use serial sampling (klinotaxis) or bilateral sampling (tropotaxis) of the concentration \cite{Rajan:2006aa}. Serial sampling depends on inter-sniff comparisons of odor concentrations between sequential sniffs that are measured at different locations. Bilateral sampling, on the other hand, depends on comparisons of odor concentrations detected by sensors located at two different positions of the body. 
\\

The ability to use inter-sensor geometry to localize odors has been observed in many animals especially insects. When one of the antennas was removed, walking fruit flies ({\em Drosophila melanogaster}) \cite{Borst:1982aa}, flying fruit flies \cite{Duistermars:2009aa}, ants ({\em Lasius fuliginosus}) \cite{Hangartner:1967aa} and honeybees ({\em Apis mellifera}) \cite{MARTIN:1965aa} showed a tendency to orient toward the intact side. Marine animals have also shown dependence on bilateral information of the odor concentration to orient. Leopard sharks \cite{Nosal:2016aa}, which are nearshore species, followed more tortuous paths and ended farther away from the shore when one of their nostrils was blocked, in contrast to control sharks which ended closer to the shore with relatively straight tracks. Crustaceans also exhibited a loss of ability to correctly orient in an odor plume and efficiently find odor sources when one of their antennules was ablated \cite{Beglane:1997aa,Derby:2001aa,Grasso:2002aa,Leonard:1994aa,Reeder:1980aa}. The detriment of loss of bilateral inputs was also shown in mammals. When one of the nostrils was partially or completely blocked, rats' accuracy in localizing odor dropped significantly and their response was biased towards the unblocked side. Their performance in tracking odor trails also declined and was less efficient\cite{Khan:2012aa,Rajan:2006aa}. Blocking a nostril in moles also biased the animal in one direction and increased the latency to find the source\cite{catania}. In this study, crossing the airflow, by inserting polyethylene tubes into the nostrils, disrupted the ability to localize sources. Likewise, human subjects' accuracy almost halved when one nostril was taped during a scent tracking task \cite{Porter:2006aa}. 
\\

Due to the behavioral and neural \cite{AON,Porter:2005aa,Rajan:2006aa} evidence of the importance of bilateral comparisons in odor localization and tracking, many have modeled animal navigation using tropotaxis \cite{Calenbuhr:1992aa,Calenbuhr:1992ab,Huang:2017aa}. A number of studies use Braitenberg vehicles (robots with simple sensor-motor connections that produce complex behaviors) equipped with bilateral sensors to detect chemicals in the environment, such as gas leaks (reviewed in section 6 of \cite{2008robot}). 
\\

In this paper, we present a mathematical analysis of tropotaxis in the presence of smooth odor sources and trails. We provide a fairly comprehensive analysis of the model dynamics, which in several cases reduces to a planar dynamical system.  In the first section, we study the dynamics on an infinite trail.  We show that there are always two stable fixed points and that there is an optimal sensor angle for attraction to the trail. We also show that the basin of attraction can be estimated from an associated integrable system.  We next consider circularly symmetric trails which include a single spot as well as circular trails.  The dynamics is more complicated there and we explore several different regimes including long sensors and sensors that are oriented behind the animal.  Finally, we consider more complicated odor landscapes such as partial trails and multiple odor sources. Here we also study trails with gaps and trails that branch and make sharp turns. 

\begin{figure}[htbp]
\centering
\includegraphics[width=3in]{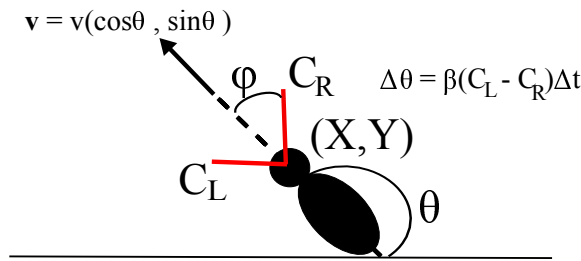}
\caption{The bilateral model: an animal centered at $(X,Y)$, heading in the direction, $\theta$. The sensors are length, $l$ with angle $\pm\phi$ around the axis of the body. Orientation is governed by the difference in concentrations at the two sensors, $C_L-C_R$ and speed is constant, $v$. }
\label{fig:animal}
\end{figure}

\section{The Model}
\label{sec:model}

The model that we will analyze describes a navigation mechanism in which the angle of the heading ($\theta$) of the individual depends on the difference between the concentration detected by the left and right sensors (See Fig. \ref{fig:animal}). The $(X,Y)$ position of the individual is a function of the heading angle and the individual's speed $v$, which we will fix to be constant: the individual is always moving. The sensors have length $l$ and are separated by an angle $\phi$ between them. They are located at the left and the right of the individual's body at positions $(X+l\cos(\theta+\phi), Y+l\sin(\theta+\phi))$ and $(X+l\cos(\theta-\phi),Y+l\sin(\theta-\phi))$  and detect odor concentration $C_L$ and $C_R$ where the concentration is generally a smooth gradient in some shape such as a line or a point source.
The bilateral olfactory navigation model equations are
\begin{align*}
 \dot{X}&=v\cos\theta    \\[-1ex]
\dot{Y}&=v\sin\theta\\[-1ex]
\dot{\theta}&=\beta \big[C_L(X,Y,\theta) - C_R(X,Y,\theta)\big].
\end{align*}
The parameter $\beta$ is the sensitivity to odor differences.  If the concentration is greater on the left, the individual turns left and {\em vice versa.} 
To make the model dimensionless, we propose a change of variables $(X,Y,t) \rightarrow (\sigma x, \sigma y, \dfrac{\sigma}{v} \hat{t})$ where $\sigma$ is the spread of concentration and $v$ is the velocity. \cmmnt{ $X=\sigma x$, $Y=\sigma {y}$ and $t=\dfrac{\sigma}{v}\hat{t}$}This will change the left sensor position to $({x}+\hat{l}\cos(\theta+\phi), {y}+\hat{l}\sin(\theta+\phi))$, the right sensor position to $({x}+\hat{l}\cos(\theta-\phi),{y}+\hat{l}\sin(\theta-\phi))$, the sensor length to $\hat{l}=\dfrac{l}{\sigma}$, and the sensitivity to concentration difference to $\hat{\beta} = \dfrac{\sigma}{v}\beta$\cmmnt{, and $\hat{r_0}=\dfrac{r_0}{\sigma}$}. The new model equations are

\begin{align}\label{eq:xyt}
\begin{aligned}
\dot{x}= \dfrac{\partial {x}}{\partial \hat{t}} &= \cos \theta \\
\dot{y}= \dfrac{\partial {y}}{\partial \hat{t}} &= \sin \theta \\
\dot{\theta}= \dfrac{\partial \theta}{\partial \hat{t}} &= \hat{\beta} \big[C_{L}({x},{y})-C_{R}({x},{y})\big].
\end{aligned}
\end{align}
These equations together with the initial conditions give us the bilateral model. We will use this dimensionless model throughout the paper unless otherwise mentioned and we will drop the \^{} for easier notation.

\section{Infinite Line}
\label{sec:infline}

\begin{figure}[htbp] 
 \centering
\includegraphics[width=5in]{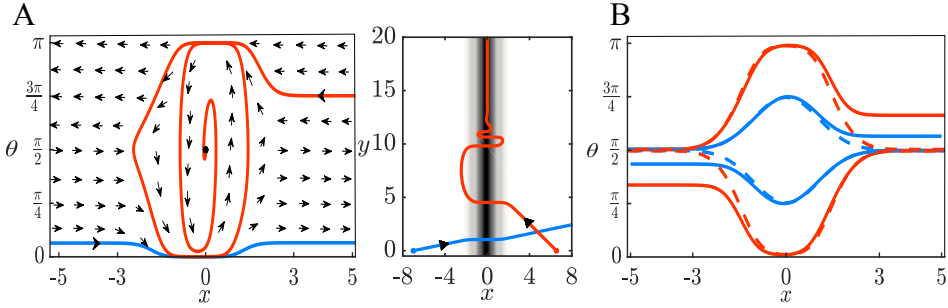}
\caption{\textbf{\label{fig:iline}}  (A) (left) Phase plane when trail is an infinite line. One trajectory converges to the stable fixed point at $(0,\dfrac{\pi}{2})$ but another does not. From the vector field, a separatrix can be noticed around the line, $\theta=\pi/2.$ (right) Projection of the solutions in the $(x,y)-$plane. (B) Basin of attraction of the trail. The dashed lines are the separatrices for the integrable system that separate the bounded solutions from the unbounded. The solid lines are the numerically simulated basins.  The blue lines represent the basin when $\beta=1$ and the red lines when $\beta = 10$ Here $\phi=1,l=0.2$.}
\end{figure}

We will start by analyzing how the model performs when the odor is along an infinite line. This corresponds to a straight trail along the $y-$axis. Here, the object is for the individual to find the trail (i.e., navigate to it) and then keep on it.  The odor concentration has a Gaussian profile and is equal to $C(x) = \exp(-x^2)$. (This is the simplification of a point source odor profile; one can use a more principled model based on advection-diffusion equation, c.f. \cite{infotaxis} Eq. 6, supplement, but the Gaussian has the advantage of being smooth at the origin making the analysis possible. Results for other odor profiles are qualitatively similar.)  Since the concentration is independent of $y$, the equations are reduced to a simple planar ODE:
\begin{align*}
\dot{x}&=\cos\theta \\
\dot{\theta}&={\beta} \big[C_{L}(x) - C_{R}(x)\big].
\end{align*}
The fixed points of the system are at $(0,\pm \dfrac{\pi}{2})$. They correspond to finding the trail and either going up ($+\pi/2$) or down ($-\pi/2$)  the trail. Here, we will limit our domain to $\theta \in [0,\pi]$, and thus the fixed point is at $(0,\dfrac{\pi}{2})$. This fixed point is stable as long as $\phi\in(0,\pi/2)$, as is the corresponding fixed point at $-\pi/2$. The trace and determinant of the linearization are respectively:
\begin{eqnarray*}
 \mbox{Tr} &=& -2\beta l^2 \sin(2\phi)\exp(-(l\sin\phi)^2) \\
  \mbox{Det} &=& 4\beta l \sin(\phi)\exp(-(l\sin\phi)^2).
\end{eqnarray*}
Since the trace is negative and the determinant is positive for all $\phi\in(0,\pi/2)$, the fixed point is asymptotically stable. This fixed point shifts horizontally if the length of the sensors are not the same. It becomes a saddle point when the sensors are crossed and then the individual will not be able to navigate the odor trail. Also, the individual will not find the trail when one of its sensors is cut. 
The left panel of Fig. \ref{fig:iline}A shows a pair of trajectories, one of which misses the equilibrium and travels off to the right and another that eventually lands on the fixed point suggesting that there is a basin of attraction for the fixed point. The right panel of Fig. \ref{fig:iline}A shows the projection of these trajectories in  $(x,y)-$plane. 
Fig. \ref{fig:iline}B shows the basin of attraction for $l=0.2,\phi=1,\beta=10,1$ in solid red and blue respectively.  (These curves are computed by integrating backwards starting at $x=\pm 5$ and $\theta$ close to $\pi/2$.) Any initial data contained within the solid curves will be attracted to the fixed point $(0,+\pi/2)$ \ and any initial data outside this will go off to $\pm\infty$. As would be expected, the blue region lies entirely in the red region.  Intuitively, if the individual is too far away from the source, unless it is nearly aligned with the trail, the concentration difference will never get large enough to allow it to correct.  We can put this intuition on a more rigorous footing by assuming the sensor length, $l$, is small to get (via Taylor's theorem): 
\[
C_L-C_R = [4l\sin\phi] \: x \exp(-x^2)\sin\theta + O(l^2)
\] 
so that we obtain an approximate system:
\begin{align*}
\dot{x}&=\cos\theta \\
\dot{\theta}&= [\beta l \sin\phi] \; x\exp(-x^2)\sin\theta.
\end{align*}
This ODE is integrable, with 
\[
E(x,\theta) := -2\beta l \sin\phi\exp(-x^2) - \log(|\sin\theta|)=\hbox{constant}.
\]
$E(x,\theta)=0$ corresponds to a pair of trajectories (shown by the dashed lines in figure \ref{fig:iline}B) that separate bounded ($E<0$) from unbounded ($E>0$) trajectories. As can be seen in the figure, these curves are reasonable approximations to the full basin of attraction (at least for $l$ small). 

\subsection{Sensor angles}
\label{sec:sa}

\begin{figure}[htbp]
\centering
\includegraphics[width=5in]{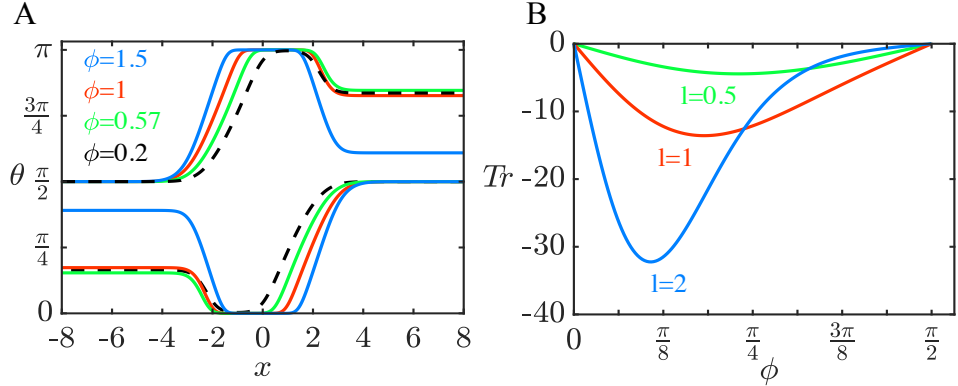}
\caption{(A) Basin of attraction for the stable fixed point $(0,\pi/2)$ for trail following as a function of initial orientation and $x-$position for 4 different sensor angles, $\phi$. Remaining parameters are $l=1,\beta=10$. (B) Trace of the linearization about the stable fixed point as the angle between the sensors varies.}
\label{fig:basphi}
\end{figure}

The sensor angles play an important role in the ability to find and follow a trail. Furthermore, they are something that can be under control of the animal, whereas sensitivity and sensor length would be difficult to vary. Fig. \ref{fig:basphi}A shows the basin of attraction for a trail with $\beta=10,l=1$ as $\phi$ is varied from the nominal value, $\phi=1$ to $\phi=0.2,1.5$ and $\phi=0.57$ (the angle at which the trace is minimum for $l=1$). Consider the upper part of the diagram (the bottom is similar under the transformation, $x\to-x,\theta\to\pi-\theta$). As $\phi$ increases toward $\pi/2$ (blue curve)  and $x(0)>0$, the individual must be more closely aligned with the trail ($\theta(0)$ closer to $\pi/2$). For $x(0)<0$, the initial heading does not matter as long as $x(0)$ is close enough to the trail and in this case, there is a slight advantage to increasing the angle. On the other hand, with small $\phi$ (black curve), there seems to be no difference from $\phi=1$ for $x>0$, but for $x<0$ the basin is decreased.  While we have not measured the precise area of the basin, it would appear that $\phi=1$ (green) has the largest; losing a little for $x<0$ but keeping the maximal amount for $x>0$. We also note that when $\phi=0.57$ (red), the basin is very close to that of $\phi=1$.      

The basin is impossible to compute analytically, but a plausible surrogate is the divergence of the vector field at the fixed point, $(x,\theta)=(0,\pi/2)$. We thus consider the trace of the linearization around the fixed point which was given above.
We plot this quantity as a function of $\phi$ for several different values of $l$ as shown in Fig. \ref{fig:basphi}B.  Clearly as $l$ increases the minimum shifts toward lower values of $\phi$. With a little bit of calculus and algebra, we find that
\[
\cos\phi_{min} = \sqrt {{\frac {{l}^{2}+\sqrt {{l}^{4}+1}-1}{2{l}^{2}}}}.
\]
The right hand side ranges between $1/\sqrt{2}$ and 1 as $l$ ranges between 0 and $\infty$.  This suggests that the sensors should have an angle between them that is between 0 and $\pi/2$.  The distance between the sensors is $2 l \sin\phi$, yielding the optimal distance to be:
\[
d_{opt}(l)=\sqrt {{{2\,{l}^{2}-2\,\sqrt {{l}^{4}+1}+2}}} .
\]
$d_{opt}$ saturates near $l=2$ at $\sqrt{2}$, which suggests that the optimal sensor distance for staying on a trail whose characteristic width is $\sigma$ will be $\sqrt{2}\sigma$.  

In sum, a single infinite odor trail greatly simplifies the dynamics to lie on the plane. There are only two fixed points, both always stable corresponding to moving up or down the trail.  There is an optimal angle for the sensors that maximizes the stability and decreases with the sensor length. The basin of attraction is well-approximated by a simple analytic formula for an associated integrable system.

\begin{figure}[htbp]
\centering
\includegraphics[width=1.5in]{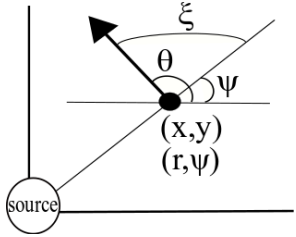}
\caption{Change of variables when odor landscape is radially symmetric. $(r,\psi)$ are the polar coordinates introduced when the individual is at position $(x,y)$ and has a heading angle $\theta$. $\xi$ is the relative coordinate where $\xi = \theta - \psi$.}
\label{fig:changeofvariables}
\end{figure}

\section{Radially Symmetric Landscapes} 
\label{sec:radial}
We now turn our attention to odor landscapes that are radially symmetric, which include point sources and circular trails. This symmetry allows us to again reduce the three-dimensional dynamical system to a planar system.
We introduce polar coordinates, $r,\psi$ ($x=r\cos\psi,y=r\sin\psi$) and the relative coordinate, $\xi=\theta-\psi$. Note that $\xi=0$ (respectively $\xi=\pi$) corresponds to heading away from (resp. toward) the source along a radial line (See Fig. \ref{fig:changeofvariables}). With these coordinates, we again obtain a planar system:
\begin{align}\label{eq:rad}
\begin{aligned}
\dot{r}&=\cos\xi\\
\dot{\xi}&={\beta} \big[C_L(r,\xi) - C_R(r,\xi)\big]-\dfrac{1}{r}\sin\xi := G(r,\xi).
\end{aligned}
\end{align}
With a radially symmetric concentration, $C(r)$, the left and right concentrations are 
\begin{align*}
C_L(r,\xi) &= C\big(\sqrt{r^2+{l}^2+2{l}r\cos(\xi+\phi)}\big)\\
C_R(r,\xi) &=C\big(\sqrt{r^2+{l}^2+2{l}r\cos(\xi-\phi)}\big).
\end{align*}
Any equilibria will have $\xi=\pm \pi/2$ and $r=\bar{r}$ chosen to solve $G(\bar{r},\pm \pi/2)=0.$  These fixed points correspond to the individual moving counter clockwise (resp. clockwise) around the source at a constant velocity. Whether such fixed points exist and whether they are stable is the subject of the rest of this section. 
  
Henceforth, we will assume the concentration has the form: $C(r) = \exp(-(r-r_0)^2)$ where the peak concentration forms a ring of radius $r_0$ around a central point. Note that $r_0=0$ is a point source. As noted above, there are two different values of $\xi$ corresponding to equilibria; since they just represent the individual going clockwise or counter-clockwise, we will focus on the latter, $\bar{\xi}=\pi/2$. 

\begin{remark} \rm  
We have chosen a simplistic model for the circular trail, $C(r,r_0)=\exp(-(r-r_0)^2)$ which is not a physical possibility. Rather, the correct form is to convolve the Gaussian with a Dirac distribution on a circle.  The result of this is:
\[
C_{real}(r,r_0) = N(r_0)I_0(2r_0r)\exp(-2r_0r)\exp(-(r-r_0)^2),
\]
where $I_0$ is the modified Bessel function of the first kind and $N(r_0)$ is chosen so that $C_{real}(r,r_0)$ has a maximum value of 1. One problem is the computation of  $N(r_0)$ since there is no simple analytical expression for the value of $r$ maximizing $C_{real}$.  For $r_0$ close to zero, the two forms are indistinguishable and for $r_0>2$, they are also quite close.  Thus it is only for values of $r_0$ around 1 that there are differences. (Recall, that we have scaled the width of the Gaussian to be 1.) We have reproduced all the phase-portraits except those in Fig. \ref{fig:phidep} using the physically correct concentration. However, we also note that we have only approximated $N(r_0)$ as no analytic expression exists and the behavior in figure \ref{fig:phidep} occurs for a very limited range of $r_0$. 
\end{remark}  

\begin{figure}[htbp]
\centering
\includegraphics[width=5in]{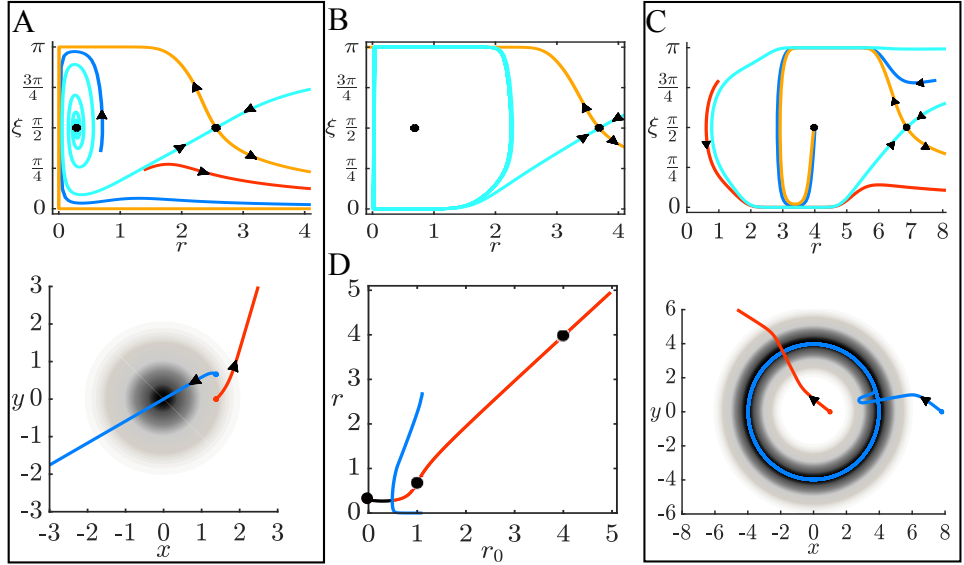}
\caption{A. (top) Phaseplane for equation (\ref{eq:rad}) for $r_0=0$, a spot  source showing an unstable spiral (near $r=0.4$) and a saddle (near $r=2.5$) along with its stable (cyan) and unstable (orange) manifolds and two trajectories. (bottom) Projection of the solutions in the $(x,y)-$plane. (B) Phaseplane for $r_0=1$. The stable manifold forms an unstable limit cycle as shown in the bifurcation diagram, D. The  fixed point inside is stable. C. (top) Phaseplane for $r_0=4$ with the same conventions as in panel A. Note the unstable spiral has become an attractor. (bottom) Projection in the $(x,y)-$plane. (D) Bifurcation diagram as a function of the trail radius, $r_0$; stable (unstable) fixed points are red (black) and unstable limit cycles are blue. Black dots correspond to $r_0=0,1,4$ and the phaseplanes in A,B,C.  Parameters are $\beta=10,\phi=1,l=1.$}
\label{fig:radial1}
\end{figure}

Fig. \ref{fig:radial1}A shows the behavior of the model when $r_0=0$, a point source.  The top shows the phase-plane for (\ref{eq:rad}). There are two fixed points, the one closest to $r=0$ is an unstable source and the larger one is a saddle point. The stable (cyan) and unstable (orange) manifolds are drawn.  While there are no attractors in this case, the stable manifolds still play an important role in the dynamics. If the initial data lies above them, then solutions in the $(x,y,\theta)$ system will pass through the odor spot as seen in the $(x,y)-$projection in the bottom of the panel. Initial data below the manifolds will veer off without getting closer to the spot. While there are no attractors (there is no ``trail'' to follow), from a practical point of view, any initial condition above the stable manifolds will ``find'' the spot. The bifurcation diagram in Fig. \ref{fig:radial1}D shows the behavior of the small $r$ fixed point as $r_0$ increases. At $r_0\approx 0.5$, the unstable source becomes a stable sink via a Hopf bifurcation. A branch of unstable periodic orbits (blue curves) emerges and terminates at an orbit homoclinic with the saddle point (not shown). We remark that for large $r_0$, the stable equilibrium is $r\approx r_0$, so the individual is centered on the trail just as in the line trail. Fig. \ref{fig:radial1}C top (bottom) panel shows the $(r,\xi)-$phaseplane ($(x,y)$ projection) for $r_0=4$. In this case, the stable manifolds form the basin of attraction for the circular trail. Any initial condition starting within the basin will find and follow the trail (blue trajectories) while outside the basin will not follow it (red trajectories).  Fig. \ref{fig:radial1}B shows the $(r,\xi)-$phaseplane for $r_0=1$.  In this case, the basin is the unstable periodic orbit that is the $\alpha$-limit set of one of the branches of the stable manifold. If one of the sensors is cut, the individual converges to a new stable periodic orbit (in $(x,y)$ plane) with a smaller radius as long as it is starting in the region bounded by the circular trail.   

\subsection{Dependence on the model parameters}
\label{sec:param}

The stabilization of the fixed point as $r_0$ increases occurs via a Hopf bifurcation. In the next sections, we explore this dependence in detail. 

\subsubsection{Sensor angle}  
\label{sec:radialsa}

\begin{figure}[htbp]
\centering
\includegraphics[width=5in]{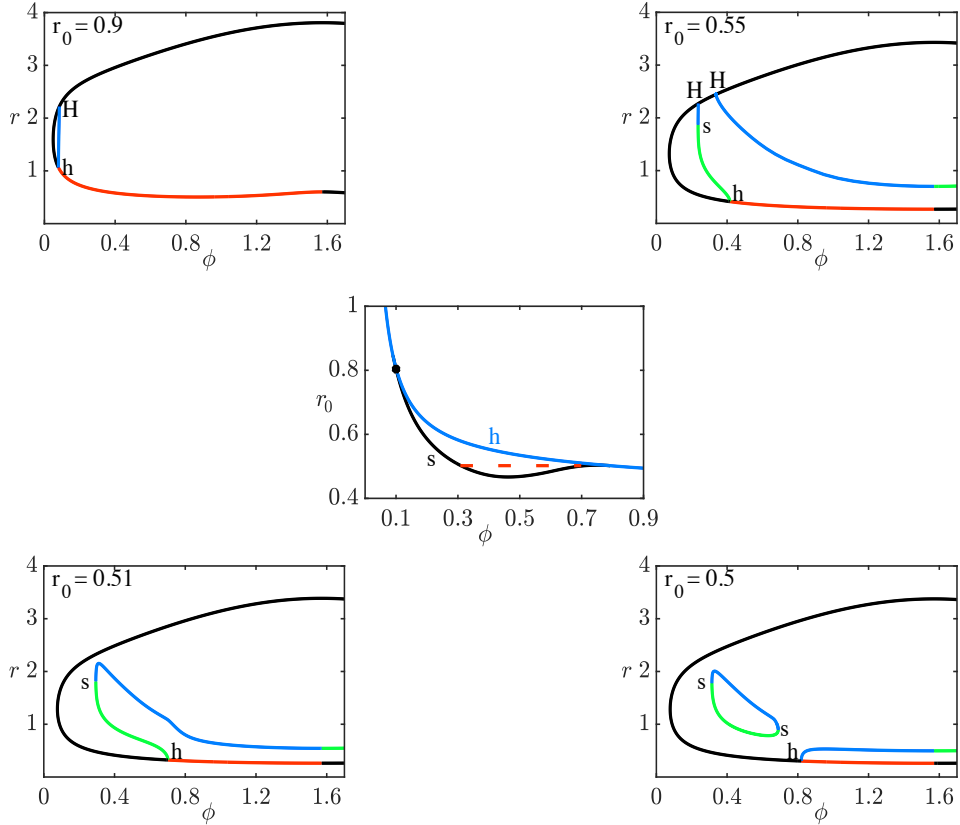}
\caption{Behavior as $r_0,\phi$ vary.  Center figure shows the two-parameter $(\phi,r_0)$ plane. Blue line denotes the curve of Hopf bifurcations. Above this curve there is a stable fixed point. The black line is the curve of saddle-node bifurcations of periodic orbits. Below the red dashed line there are 2 saddle-node of limit cycles (isola).  One-parameter bifurcation diagrams are shown for different values of $r_0$ as $\phi$ varies.  Black (Red):unstable (Stable) fixed points; Blue (Green): unstable (stable) periodic orbits. (h), Hopf bifurcations; (s), saddle-node of limit cycles; (H), saddle-homoclinic orbits.}
\label{fig:phidep}
\end{figure}

The sensor angle, $\phi$ provides an interesting picture. We first note that if we let $\hat{\phi}=\pi-\phi$ and $\hat{\xi}=\xi+\pi$ then equation (\ref{eq:rad}) becomes:
\begin{eqnarray*}
\frac{dr}{dt} &=& -\cos\hat{\xi} \\
\frac{d\hat{\xi}}{dt} &=& -\left(\beta\big[C_L(r,\hat{\xi})-C_R(r,\hat{\xi})\big] -\sin\hat{\xi}/r\right),
\end{eqnarray*}
with $\hat{\phi}$ replacing $\phi$.  Angles $\phi\in (\pi/2,\pi)$ correspond to the individual having its sensors behind it. This calculation  shows that the vector field for $\phi\in(\pi/2,\pi)$  is the same as that for $\phi\in(0,\pi/2)$ in reverse time. 
Thus, for example, unstable periodic orbits for $\phi\in(0,\pi/2)$ become stable periodic orbits for $\phi\in(\pi/2,\pi).$   Additionally, note that when $\phi=\pi/2$, then Eq.(\ref{eq:rad}) is a reversible system, since $\xi\to\xi+\pi$ takes $t\to-t$.  Thus, for fixed $r_0$ and increasing $\phi$ from 0, there will be  three Hopf bifurcations; the middle one is degenerate and is at $\phi=\pi/2$, the reversible system.  To get more insight into the full dynamics, we look at the $(\phi,r_0)$ parameter plane in more detail.  Fig. \ref{fig:phidep} shows bifurcation diagrams as $\phi$ varies for several different values of $r_0$.  There are several notable features.  The central diagram shows the curves of Hopf bifurcations (blue) in addition to curves of saddle-nodes of limit cycles (SNLCs, black).  The latter curve is non-monotonic, so that there is a region (below the red dashed curve), where there can be two SNLCs.  The lower right diagram shows that these delineate an isola (isolated branch) of periodic orbits. As $r_0$ increases, this isola merges with the branch of unstable periodic orbits (lower left diagram). Between $r_0=0.51$ and $r_0=0.55$, the stable and unstable branches collide with the saddle at a saddle-homoclinic bifurcation (shown as H in the upper right diagram). Finally, the SNLC merges with the Hopf bifurcation curves (shown by the asterisk in the central figure) leaving an unstable periodic orbit (upper left diagram; the other unstable periodic orbit is not shown). The apparent existence of stable periodic orbits for small radii trails and small sensor angles implies that there is a stable torus in the full $(x,y,\theta)$ system.      

\subsubsection{ Sensor length} 
\label{sec:sl}

\begin{figure}[htbp]
\centering
\includegraphics[width=5in]{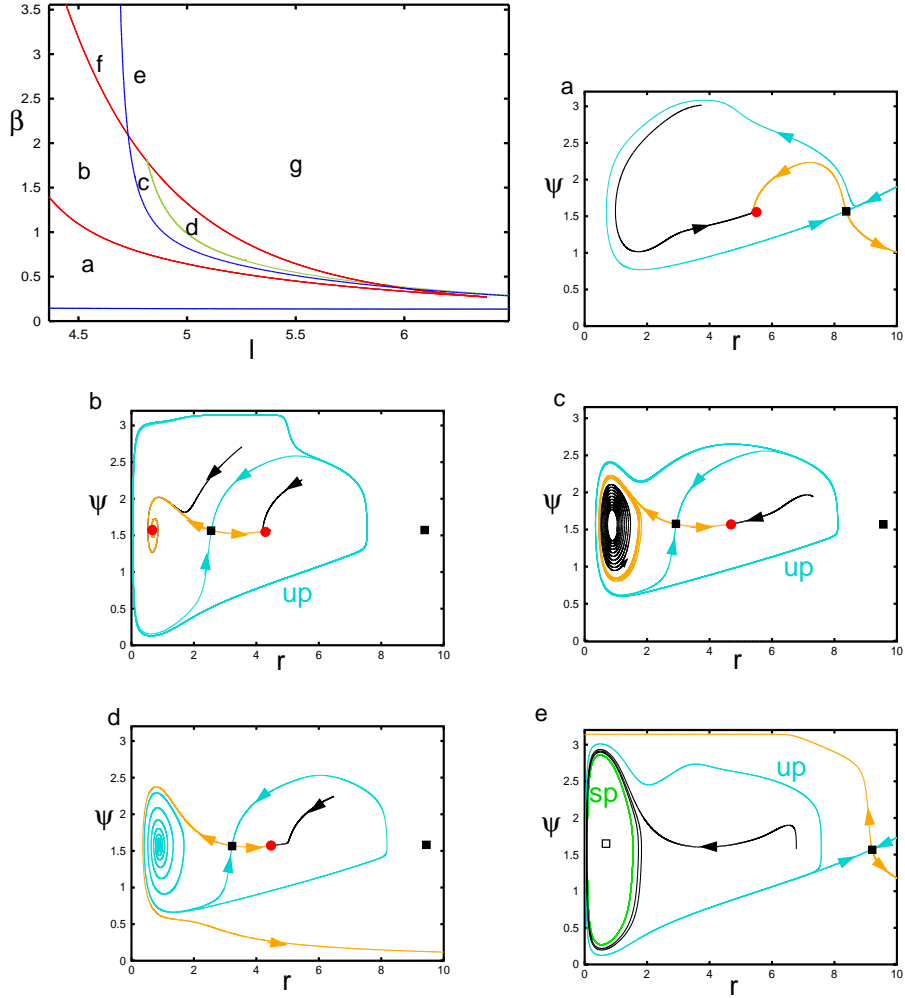}
\caption{Dynamics on circular trail (here $r_0=4,\phi=1$) when $l$ is large and $\beta$ is small.  The dynamics is organized by the saddle-nodes or folds of equilibria (red), the Hopf bifurcation (blue), and a homoclinic bifurcation (olive).  Phaseplanes in the representative regions are depicted.  Stable (cyan) and unstable (orange) manifolds of the saddles (filled black squares) are shown along with some representative trajectories (black). Stable fixed points are red circles, saddles are black squares, unstable nodes are hollow squares. UP:unstable periodic orbit; SP:stable periodic orbit.  Region f is like region e, but the stable periodic orbit is replaced by a stable fixed point. In region g, there are no attractors. Panel e shows a stable isolated limit cycle in green. More details in the text. Parameters $(l,\beta)$: (a) (4.5,0.5), (b) (4.5,2), (c) (4.85,1.25), (d) (4.93,1.25), (e) (4.72,3)  }
\label{fig:biga}
\end{figure}

\begin{figure}[htbp]
\centering
\includegraphics[width=3in]{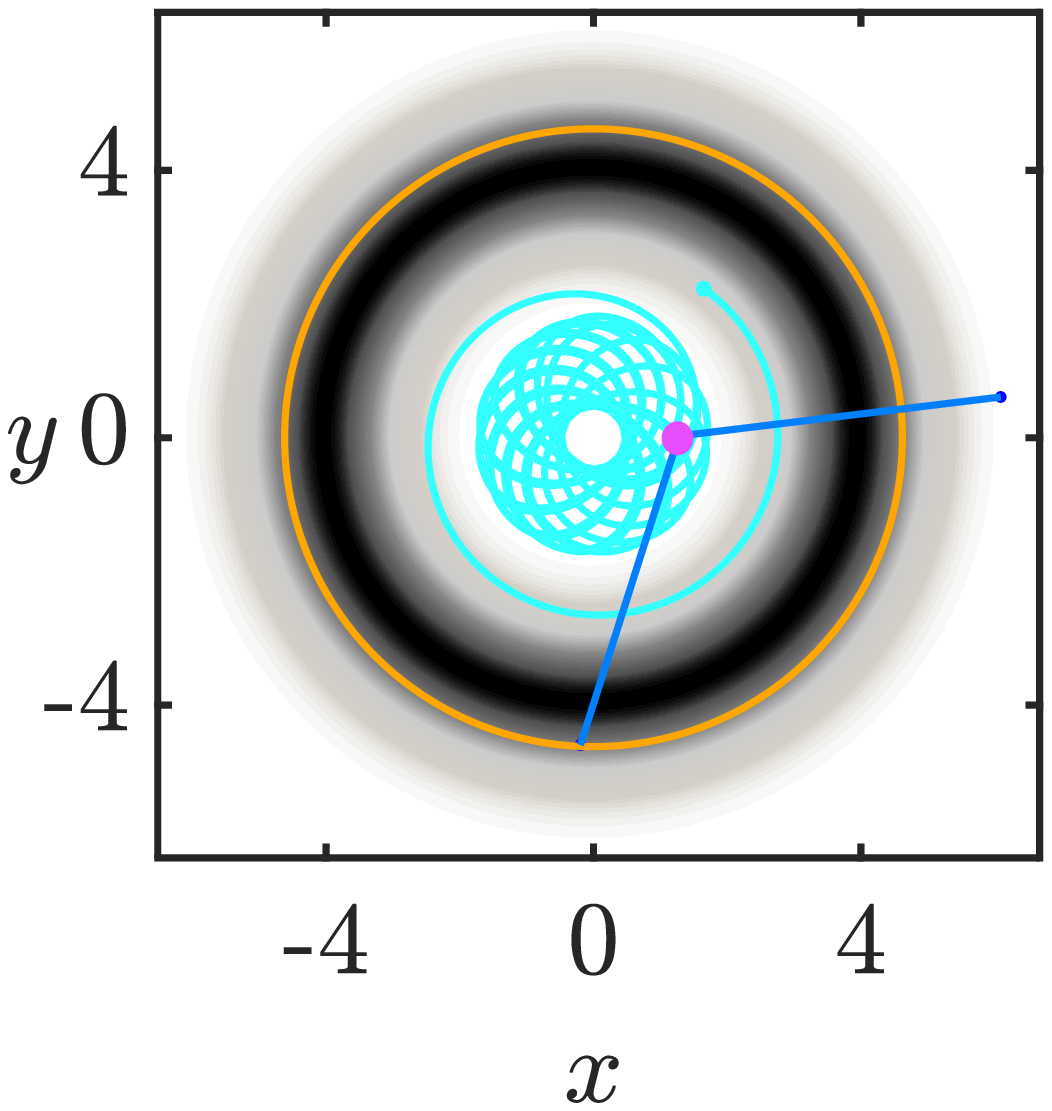}
\caption{Projection of the trajectory of the individual in the $(x,y,\theta)$ model in region c of Fig. \ref{fig:biga}. Outer orange circle is a stable path of the individual, grayscale shows trail concentration. Stable torus solution shown in cyan. Magenta spot is the individual with the sensors drawn to scale in blue. Animation can be found in supplementary video. }
\label{fig:bigatraj}
\end{figure}

Surprisingly, we have found multistability on circular trails of radius, $r_0$, for sensors that have the same approximate length $l\approx r_0$ and small attraction, $\beta$. 
Figure \ref{fig:biga} shows some examples of the dynamics. Here, we choose $r_0=4$ and $l$ between 4 and 6, while letting $\beta$ range between 0.5 and 3.5.  The dynamics is organized around the two parameter curves of various bifurcations (not all of them are shown, either for clarity or for inability to follow them).  In the figure, curves of saddle-node bifurcations of equilibria (SNE) are shown in red, Hopf bifurcations in blue, and a homoclinic bifurcation in olive.  Phaseplanes in some of the regions are shown. We emphasize once again, that stable fixed points (limit cycles) in this reduced system correspond to stable periodic orbits (tori) in the full three-dimensional model (See Fig. \ref{fig:bigatraj}). Starting in region (a), there is a single attracting fixed point whose basin is delineated by the stable manifolds of the outer saddle. (As we will eventually encounter another saddle point, the outer one will be the one that is at roughly $r=9$.  It persists throughout the figure.) Two bifurcations occur as we move from a to b. First, there is a homoclinic bifurcation at the outer saddle leading to an unstable periodic orbit (UP) that plays the role of the basin for the fixed point. (This is not shown as a separate phaseplane since  the attractor structure is still the same.) As we cross the red curve into region b, two new fixed points arise: a stable node and a saddle. The  UP continues to provide the basin, but the stable manifolds of the inner saddle (near $r=2$) split this basin between the two stable fixed points. Recalling that $r_0=4$, we see the outer fixed point shows the individual following the trail while with the inner stable fixed point the individual makes smaller circles within the trail.  In the transition from b to c, the inner fixed point undergoes a Hopf bifurcation and spawns a stable periodic orbit (SP). Thus, in the $(x,y,\theta)$ model there is bistability between the individual tracking the trail and a quasiperiodic trajectory that lies near the center of the trail. Fig. \ref{fig:bigatraj} shows the dynamics in the $(x,y)$-plane.  The transition from c to d occurs through a homoclinic bifurcation (olive curve) where the SP disappears.  The result is just a single attractor.  In d to g, this attractor is lost via a SNE and there remain no attractors. The path from c to e occurs via a SNE leaving just a SP whose basin is determined by the UP. The transition from e to g occurs when the SP and the UP (SNLC) merge and disappear.  The transition from e to f occurs when limit cycle disappears through a reverse Hopf bifurcation stabilizing the fixed point shown by the hollow square. Region f has only one attractor, this stabilized fixed point is near $r=1$ and is not shown.  We were unable to compute the curve of SNLCs delineating the transition from e to g.        

\subsubsection{Basins of attraction}
\label{sec:basins}

\begin{figure}[htbp]
 \centering
        \includegraphics[width=5in]{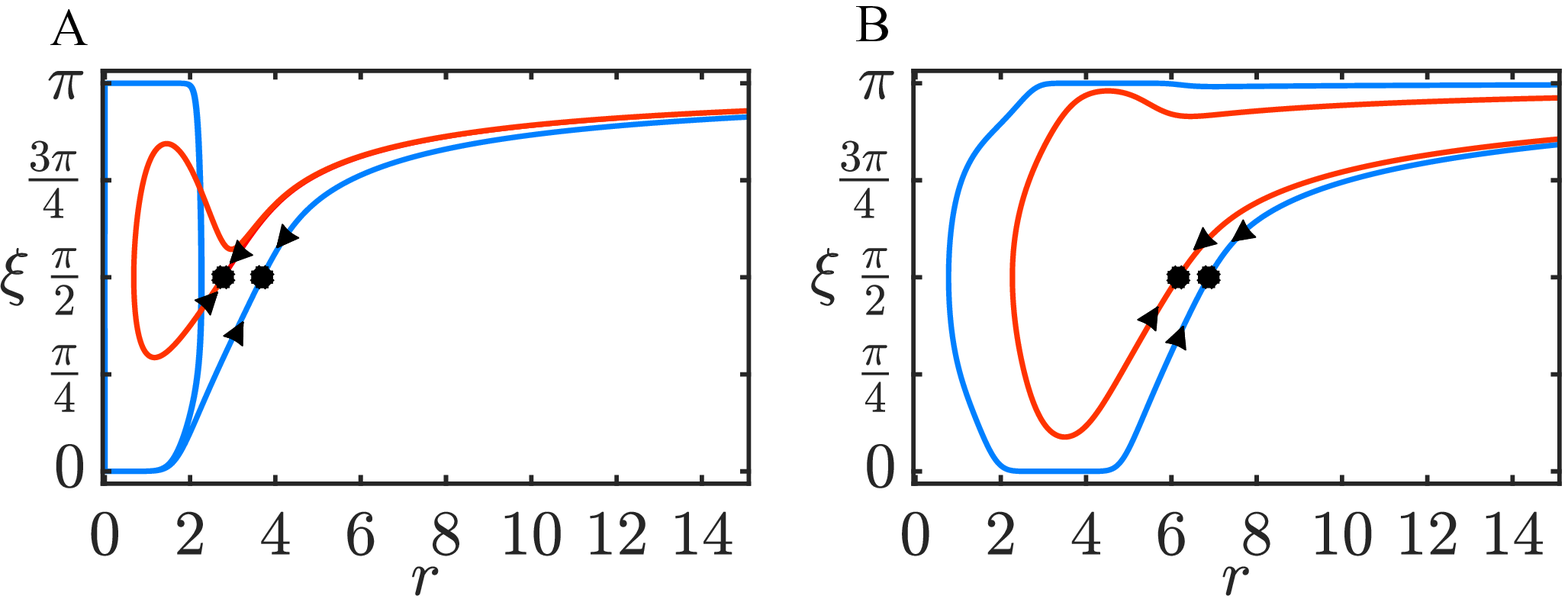}
      \caption{(A) Basin of attraction when trail is circular with radius $r_0 = 1$. (B) Basin of attraction when trail is circular with radius $r_0 = 4$. For both figures, blue and red lines correspond to the basin when $\beta =10$ and $\beta = 1$ respectively. 
\label{fig:beta}}
\end{figure}

Given a circular trail sufficiently large that there is a stable fixed point, we first examine the dependence of the basin on the radius and the turning sensitivity, $\beta$ in Fig. \ref{fig:beta}. In Fig. \ref{fig:beta}A, $r_0=1$ and $\beta=1,10$  while in  \ref{fig:beta}B, $r_0=4$. For smaller radii, higher sensitivity does not necessarily mean that the basin will be bigger. Indeed, there are initial conditions that lie in the basin of attraction for $\beta=1$ (red) , but not when $\beta=10$ (blue).  On the other hand for large radii (Fig.  \ref{fig:beta}B), the basin for $\beta=10$ contains that for $\beta=1$.  

Since there are no stable fixed points for spot location, we can consider the ability of an individual to orient toward a spot given that it is frozen ($v=0$) at a distance, $r$, from the spot. In this case, we have a simple one-dimensional system:
\[
\dot{\xi}={\beta} \big[C_L(r,\xi) - C_R(r,\xi)\big]
\]
with a stable fixed point at $\xi=\pi.$  The eigenvalue around this fixed point is:
\[
\lambda(r,\l,\phi)=-\beta 4lr\sin(\phi)\exp(-r^2-l^2+2lr\cos(\phi))
\]
and, as with the trail, this has a minimum at a particular value of $\phi$:
\[
\cos\phi=\frac{-1+\sqrt{16 (rl)^2+1}}{4 rl}:=M.
\]
As $rl\to0$, $M\to 0$ and as $rl\to\infty$, $M\to 1$. In particular, this suggests that close to the spot ($rl$ small), the animal should keep its sensors near $\pm \pi/2$ while keeping them close to 0 when it is far from the spot.    

\subsubsection{Integrability}
\label{sec:integ}

\begin{figure}[htbp]
\centering
\includegraphics[width=3in]{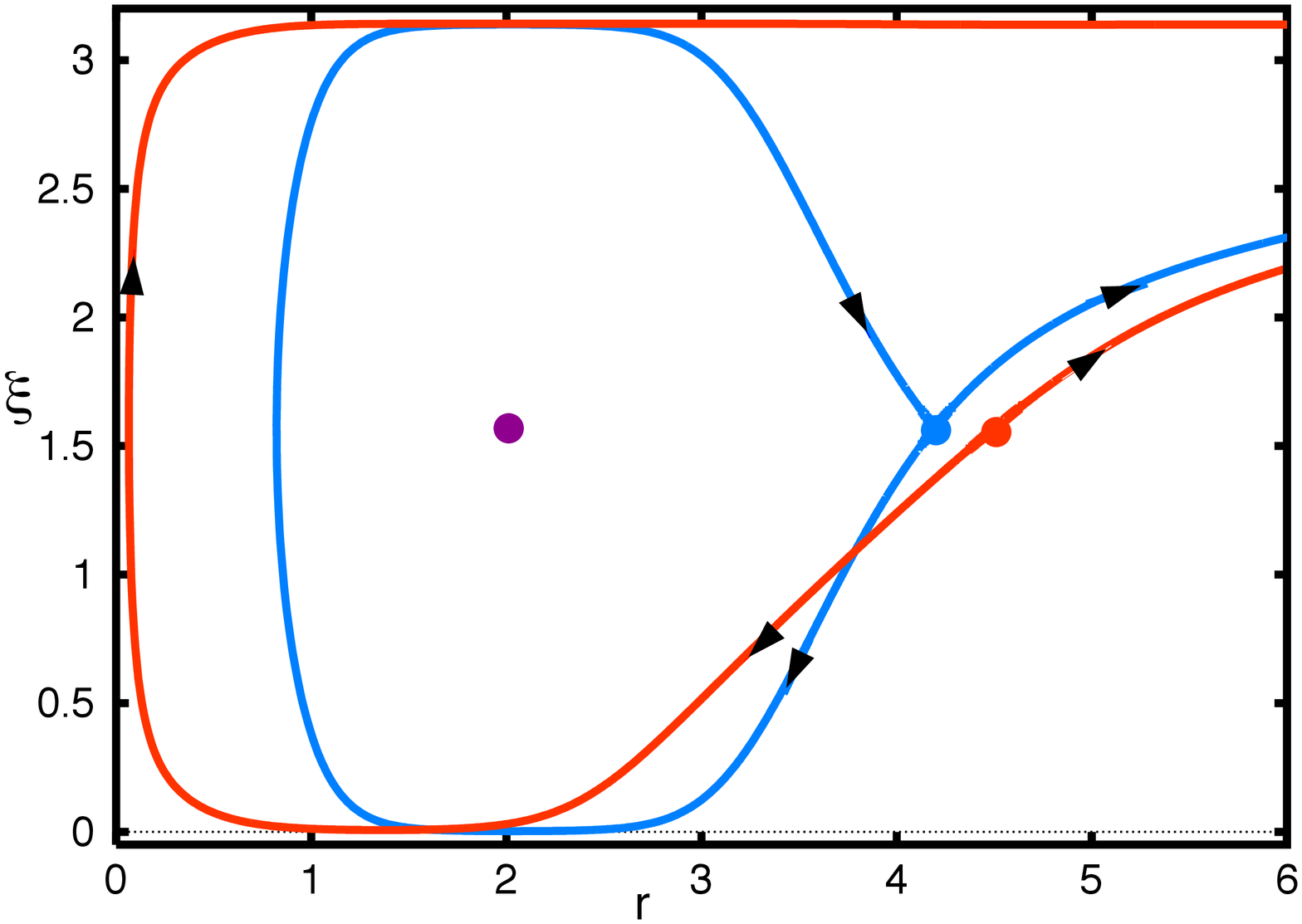}
\caption{Comparison of the basin of attraction for the full model (\ref{eq:rad}) (red) with that of the integrable approximation (\ref{eq:polin}) (blue) for $l=1,\beta=4,r_0=2,\phi=1$. Saddle points are shown in their respective colors. The stable fixed point and nonlinear center are nearly coincident and shown in purple.}
\label{fig:polin}
\end{figure}

As in the case of an infinite line, system (\ref{eq:rad}) can be approximated by an integrable system for small $l$:
\begin{eqnarray}\label{eq:polin}
\begin{aligned}
\dot{r} &= \cos\xi \\
\dot{\xi} &= [4l\beta\sin(\phi)(r-r_0)\exp({-(r-r_0)^2})-1/r]\sin\xi ,
\end{aligned}
\end{eqnarray}

with 
\[
E:= log(|\sin\xi|)+2 l \beta\sin(\phi) \exp({-(r-r_0)^2})+log(r) =\hbox{constant}.
\]
For $K:=\beta l \sin(\phi)$ large enough, the integrable system has a saddle and a nonlinear center; the stable manifolds of the saddle form a good approximation for the basin of attraction for (\ref{eq:rad}), even for $l=1$, over a wide range of the other parameters.  This calculation does not say anything about the stability of the fixed point; rather, it gives some insight into the regions of attraction. Figure \ref{fig:polin} shows that the even for $l=1$, the basins of the full equation (\ref{eq:rad}) and the integrable system (\ref{eq:polin}) are close.    \\  

As with the linear trail, radially symmetric odor gradients can also be reduced to planar dynamical systems. Nevertheless, they produce complex behavior including multi-stability and different types of stable and unstable limit cycles. Circular trails with a large enough radius lead to a stable movement clock-wise or counter-clockwise around the trail when the sensors are short.  Such trajectories are seen in so-called ant-mills (where large populations of ants move in a circular trail until they die of exhaustion)\cite{schneirla1944}. Because the individual has a constant speed, it is not possible for the point source to be an attractor. However, the model does take the individual toward the source (depending on its initial distance and heading), so, in a real situation where the source is some reward the animal would stop moving when it reached the source.  

\section{Multiple sources}
\label{sec:multsrc}

When an animal is searching for food, there can be  multiple  sources that affect the concentration detected and could be used to localize an odor source.  We next study how the bilateral model behaves in the presence of two  odor sources. With more than one source, the radial symmetry is broken and we cannot exploit the reduction in dimension used above. Thus, we will use the $(x,y,\theta)$ system and the concentration detected will be the sum of the Gaussian concentration of the spots.  

\begin{figure}[htbp]
\centering
\includegraphics[width=5in]{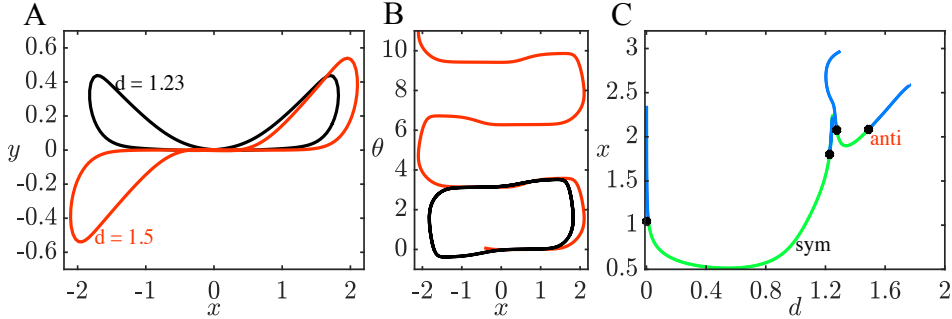}
\caption{Two different types of trajectories for concentrations with two odor sources located on the $x$-axis a distance $d$ apart, centered at $x=0$. (A) Projection into the $(x,y)$ plane; (B) Projection in the $(x,\theta)$ plane; (C) Bifurcation diagram for the two different cases in (A and B) as $d$ varies.  Other parameters are $\beta=20,l=0.5,\phi=1.$}
\label{fig:bin2as}
\end{figure}

Without loss of generality, we  place the {\em two} point sources at a distance $d$ from each other on the $x$-axis and analyze the dynamics of Eq. (\ref{eq:xyt}). 
The odor concentration at the first spot is $C_1(x,y)  = A_1\exp\big(-\big((x+d/2)^2+y^2\big)\big) $, and at the second spot is $C_2(x,y) =A_2 \exp\big(-\big((x-d/2)^2+y^2\big)\big)$ where $A_1$ and $A_2$ are positive, possibly different, amplitudes.  Thus, the concentration detected at the sensors is
\begin{align*}
C_L(x,y) &= C_1(x_L,y_L) + C_2(x_L,y_L) \\
C_R(x,y) &= C_1(x_R,y_R) + C_2(x_R,y_R),
\end{align*}
 $x_{L,R},y_{L,R}$ are as in Fig. \ref{fig:animal}.  
 
\begin{figure}[htbp]
\centering
 \includegraphics[width=5in]{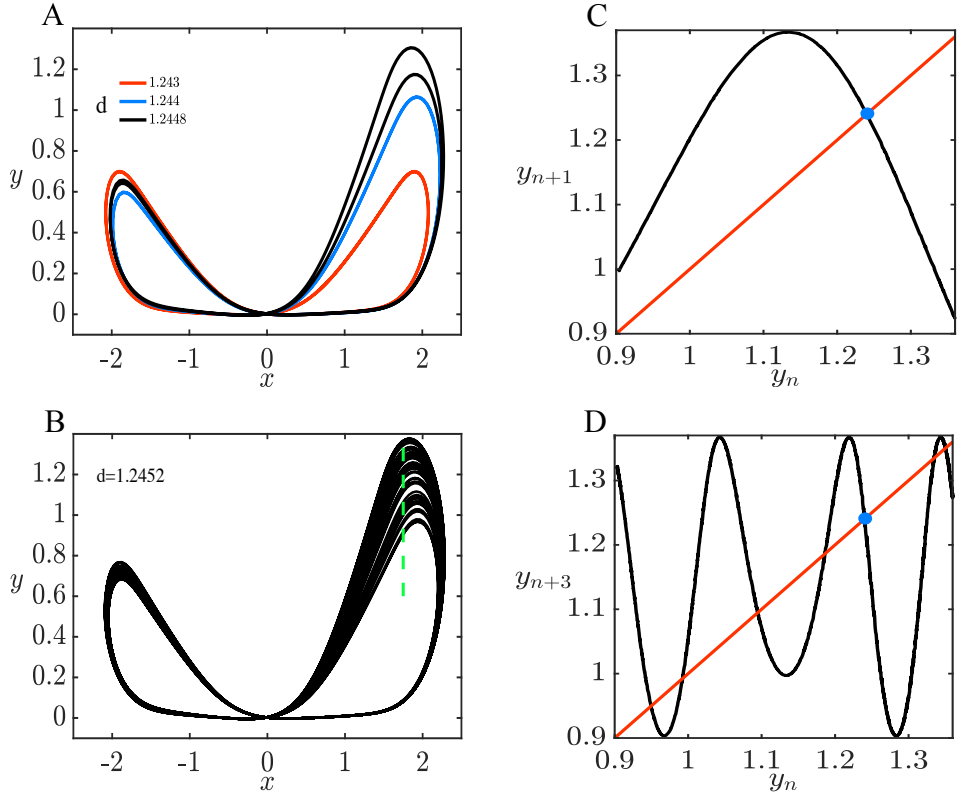}
\caption{Behavior of Eq. (\ref{eq:xyt}) when there are two Gaussian sources at $(x,y)=(\pm d/2,0).$ (A) As $d$ increases, the symmetric periodic solution (red) loses stability and gives rise to a stable asymmetric solution (blue). Increasing $d$ leads to a period doubled solution (black) which also loses stability as $d$ increases. (B) Presumably chaotic behavior for $d=1.2452$. (C) Poincare map through $x=1.75$ for the solution in (B). Blue circle is unstable periodic orbit. (D) Same Poincare section  showing the numerical existence of a period three orbit shown by the intersections of the $n+3$ iterate with the diagonal. The blue filled circle shows the period one fixed point. (Parameters are as in Fig. \ref{fig:bin2as}.)}
\label{fig:bin2chaos}
\end{figure}

Recall that in the case of a {\em circular trail}, there are stable fixed points in the polar form of the equations which correspond to circular periodic orbits in the $(x,y,\theta)$ system.  Since the individual must maintain a constant speed, we cannot expect any fixed points in the $(x,y,\theta)$ system, so we will look for periodic orbits.  We fix $\beta=20,l=0.5,\phi=1$ in this section; the default values of $\beta,l$ produce periodic orbits for a range of $d$, but the behavior is not as rich. In Figure \ref{fig:bin2as}A, we show two qualitatively different trajectories projected in the $(x,y)$ plane for spots placed a distance $d$ on the $x$-axis.  At small values of $d$ the trajectory is symmetric (black curve) and the heading, $\theta$ oscillates around $\pi/2$ (Fig. \ref{fig:bin2as}B, black) (topological winding number of 0). There is an analogous curve where $y(t)<0$ and $\theta$ oscillates about $3\pi/2$.  For a larger value of $d$, we find an anti-symmetric trajectory (Fig. \ref{fig:bin2as}A, red) and in this case, $\theta$ goes through all values with a net increase of $2\pi$ after each cycle (Fig. \ref{fig:bin2as}B, red) (topological winding number of 1).  Fig. \ref{fig:bin2as}C shows the one-parameter bifurcation diagram as $d$ changes for the symmetric and the anti-symmetric paths. The stability of these is lost at branch points marked by the filled blue circles. If we follow the symmetric branch point at the high value of $d$ (close to 1.25), then a stable branch of asymmetric solutions emerges. This is shown in Fig. \ref{fig:bin2chaos}A as the blue curve. Increasing $d$ along this asymmetric branch leads to a periodic doubling bifurcation (shown as the black curve). Further increases lead to presumably chaotic behavior, shown in Fig. \ref{fig:bin2chaos}B in the $(x,y)-$plane.  To further quantify the chaos, we take a Poincare section through $x=1.75$ and plot the points $(y_n,\theta_n)$ where $x$ crosses from right to left. We find (not shown) that these points appear to lie along a one-dimensional curve, indicating that the underlying chaos can be understood by a one-dimensional map. Fig. \ref{fig:bin2chaos}C shows the map where we plot $(y_n,y_{n+1})$. It appears to be a typical cap map. The periodic orbit (blue circle) is unstable as the slope through it is less than -1.  Fig. \ref{fig:bin2chaos}D shows $(y_n,y_{n+3})$ plotted and a clear period 3 orbit that is also unstable.  Since the underlying dynamics seems to be governed by a one-dimensional map, we believe that Fig. \ref{fig:bin2chaos}B represents a truly chaotic orbit. Additionally, the maximal Liapunov exponent is 0.045, a positive number, yet another character of chaos. 

\begin{figure}[htbp]
\centering
\includegraphics[width=5in]{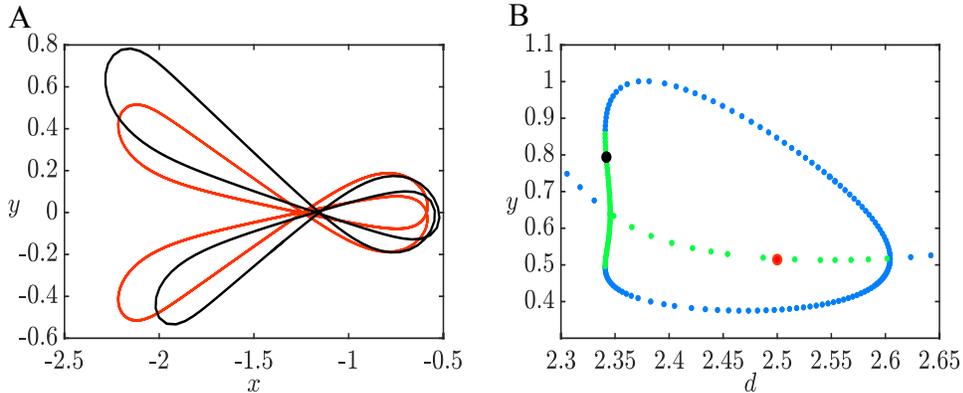}
\caption{Two distant sources. (A) Stable periodic circling around the source at $(-d/2,0)$ with the other source located at $(d/2,0)$ with $d=2.5$ (red) and $d=2.34$ (black). (B) Bifurcation of the isolated periodic orbit as $d$ changes.  There are two pitchfork bifurcations whose branches form an isolated loop. Filled circles correspond to orbits depicted in A. Remaining parameters as in Fig. \ref{fig:bin2as} } 
\label{fig:2srcdist}
\end{figure}

As the previous figures show, if the spots are close to each other, there can exist solutions where the individual circles {\em both of them}. Furthermore, when there is an isolated spot, there are no stable bounded solutions as we saw above. However, the presence of a distant spot (at least over a small range of distances) can stabilize periodic orbits around a spot. Fig. \ref{fig:2srcdist}A shows two different stable trajectories around a source at $(-d/2,0)$. The red solution is symmetric about the $y$-axis ($d=2.5$) and the black solution has lost the symmetry ($d=2.43$). This branch of periodic solutions exists for a narrow range of values of $d$ as shown in the bifurcation diagram in Fig. \ref{fig:2srcdist}B.  In particular as $d$ decreases, there is a supercritical pitchfork bifurcation that leads to the stable asymmetric solution shown in panel A. For $d$ increasing,there is a subcritical pitchfork which together with the other pitchfork forms an isolated branch of asymmetric solutions.      

\begin{figure}[htbp]
\centering
        \includegraphics[width=5in]{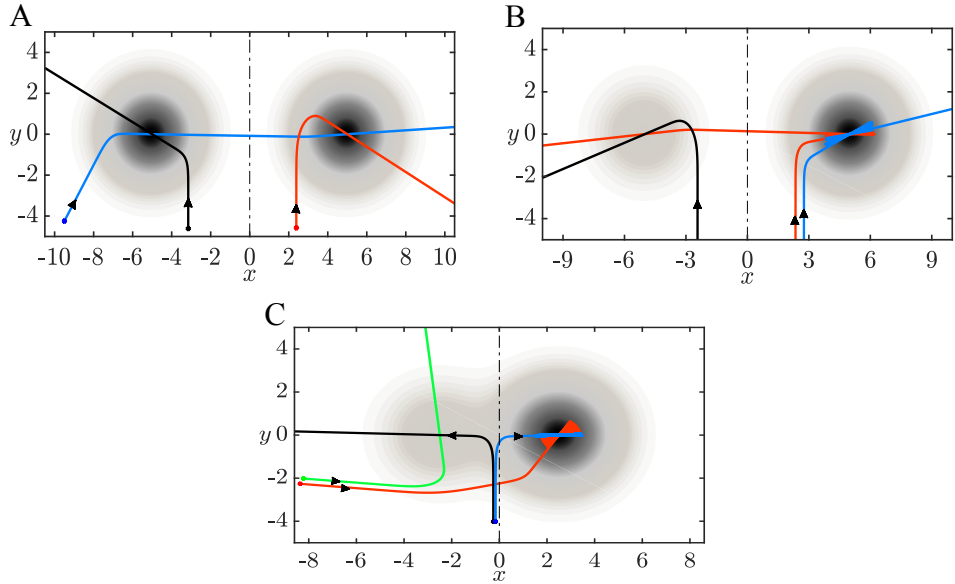} 
\caption{ Different trajectories when: (A) Both sources have the same amplitude ($A_1=A_2=1$) and are at a distance ($d=10$) where the 2 sources are distinguishable. (B) Second source has significantly larger amplitude ($A_2=5$). (C) Second source has significantly larger amplitude and the sources are closer to each other ($d=5$) Other parameters as in Fig. \ref{fig:bin2as}.} \label{fig:traj2src}
\end{figure}

Another interesting question is how the behavior changes when the concentrations at the spots are different in magnitude. 
 Fig \ref{fig:traj2src}A shows trajectories when the amplitudes of the spot are equal and the spots are at a relatively large distance from each other (such that there is no periodic orbit encircling them).  Depending on the initial position, trajectories either pass through both spots, just one of the spots or miss them both. In all cases, however, the trajectories diverge. This is also true when we increase the amplitude of one of the spots by 5-fold as in Fig \ref{fig:traj2src}B. Note that the individual spends some time wandering around the spot with higher intensity before wandering off. On the other hand, when we bring the spots closer to each other as well as increase the amplitude (Fig \ref{fig:traj2src}C), the trajectories that go to the spot with larger amplitude will oscillate around this spot. Thus, the existence of the weaker spot at a distance can stabilize the trajectory around the spot with a higher concentration, just as we saw in Fig. \ref{fig:2srcdist}.  The periodic solution shown in Fig. \ref{fig:traj2src}C persists for much larger values of $A_2$ and will also persist for $A_2$ reduced to 1, where the resulting periodic solution is the same as that seen in Fig. \ref{fig:2srcdist}A (red).  

More complex dynamics can occur with three or more sources. In this case, however, there are many different possible configurations thus we will not consider them further.

\section{Finite Trails}
\label{sec:ftrails}
      
We have looked at how the bilateral model performs when we have an infinite line and circular trails. Now we will examine its behavior on a finite line segment and a finite line segment with gaps, sharp angles and branches, as these cases can be tested in animal behavior experiments. 
 
\begin{figure}[htbp]
\centering
        \includegraphics[width=5in]{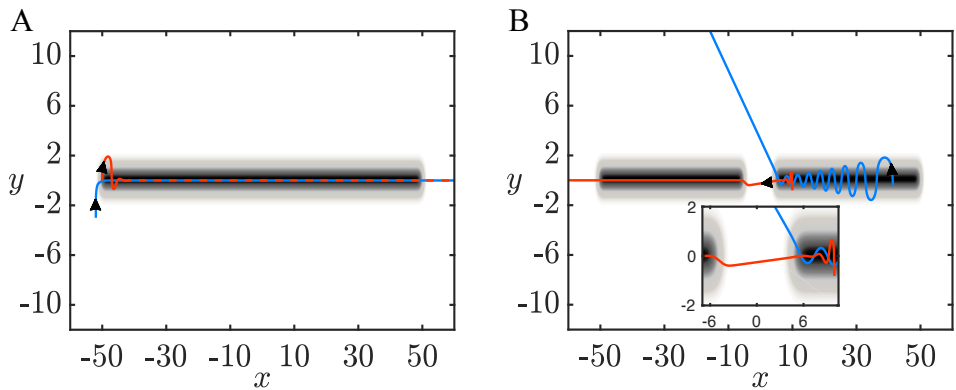} 
      \caption{\label{fig:gap} (A) Trajectories on a segment trail. Red line is when $\beta$ is 5 fold smaller than the blue line trajectory. (B) Trajectories can either cross gaps or lose the trail depending on $\beta$ or length of nares $l$. Red trajectory is when $l=0.4$ and Blue trajectory is when $l=0.1$.} 
      \end{figure}

If we start close enough to a segment trail, the model will find the trail, follow it and then leave it. When $\beta$ or $l$ is small, trajectories will have damped oscillations that decay slower as we decrease $\beta$ or $l$ (Fig \ref{fig:gap}). The starting angle affects the trajectory orientation; most trajectories continue to the right when $\theta_0$ is less than $\dfrac{\pi}{2}$  and to the left when  $\theta_0$ is larger than $\dfrac{\pi}{2}$. Similarly, if we start around the gap, then we take either the left or right branch depending on the starting position and angle. Also, we can find the trail from significantly farther distances when we start around the gap which is also the case when we start around the beginning or end of the trail. An individual will cross the gap and reacquire the trail when the gap is in a line trail that has no angles or turns. This is true because once the trail is acquired in the bilateral model, the individual will keep moving straight on it. However, if either $\beta$ or $l$ is small, and the oscillations are large near the gap, the model will sometimes lose the trail as in Fig \ref{fig:gap}B. 

\begin{figure}[htbp]
        \centering
        \includegraphics[width=5in]{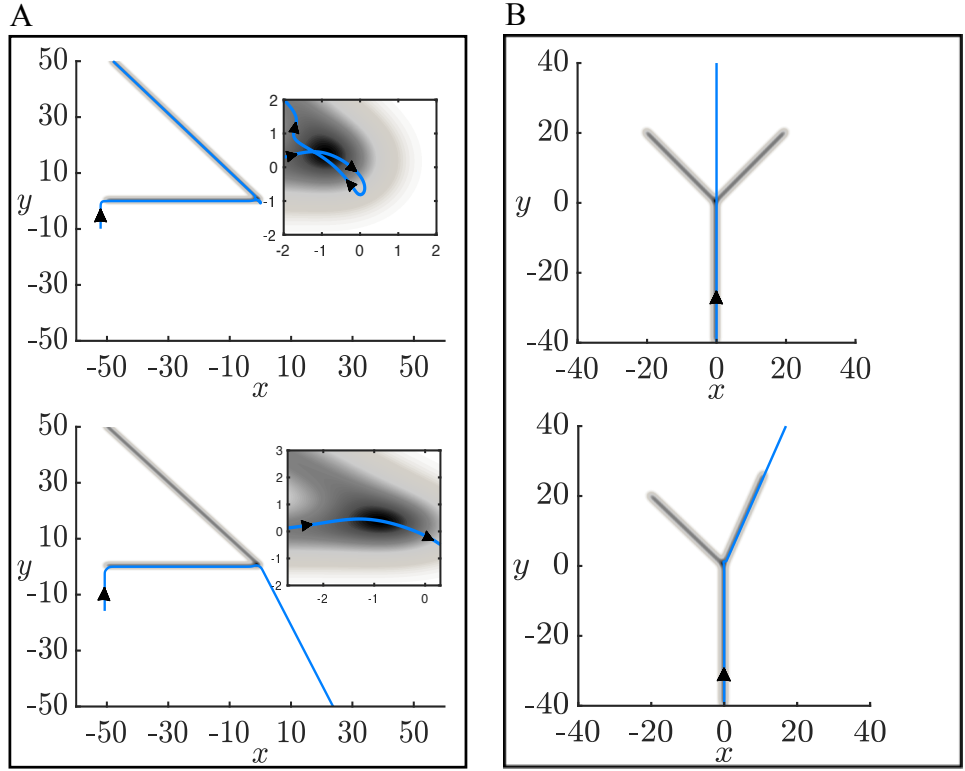} 
      \caption{ \label{fig:yang} A. (top) Trail with angle very close but greater than $ \frac{\pi}{4}$. Zoomed on how model is able to correct and find trail. (bottom) Trail with angle equal to $\frac{\pi}{4}$. Zoomed on how model can not sense the change in the angle and loses trail. B. (top) Y trail, the branches are at equal angles from the main trail. (bottom) Y trail where the branches are at different angle from the main trail. The blue line is a trajectory starting at the main trail.}
\end{figure}

If there is an angle in the trail, then it must be larger than $\dfrac{\pi}{4}$ for the model to follow it easily. In Fig \ref{fig:yang}A, the model is able to correct and follow the trail when the angle is slightly bigger than $\dfrac{\pi}{4}$, but as soon as the corner angle is $\dfrac{\pi}{4}$, the model loses the trail. 

When the trail bifurcates into two branches, the angle and amplitude of each branch determine the trajectories of the model as it passes the branch.  We observe that when an individual starts on the main trail that has two branches at equal angle and amplitude (top panel of Fig \ref{fig:yang}B), it will continue on a straight path as there is no difference in the left and right concentrations due to symmetry.  However, if the symmetry is broken, say, the branch angles are unequal,  (bottom panel of Fig \ref{fig:yang}B), then the individual will go towards the branch that requires the least amount of turning. 
This might not be the case if the concentrations on the two branches differ as the model will always turn toward the higher of the two concentrations at the sensors.
    
Trails with gaps and finite trails are similar to an infinite trail over the period of time in which the individual is on the segment.  since once the individual finds the trail, it stays on it. If the trail is short and $\sigma$ is large, then, there is behavior like two close spots, e.g.  Fig. \ref{fig:bin2as}, otherwise, the individual eventually reaches the end of the trail and moves away.  Thus, in these cases, there are no attractors and basins, bifurcations, etc do not make sense.   We have included the results on branched and finite trails mainly because they provide for the possibility of experimentally testing some of the results. Indeed, some preliminary experiments in the lab of Nathan Urban examine the paths of mice that are trained to follow trails when the trails branch and have gaps.   

\section{Discussion}
\label{sec:disc}

\cmmnt{Many animals use comparison between bilateral inputs as the fundamental strategy to locate and follow odors.  Animals that employ other strategies  can still make use of stereo sensing to increase efficiency and accuracy since the information provided by the two sensors is non-redundant \cite{JDV,Draft2018}.} In this paper, we analyzed a simplified dimensionless model that describes the use of bilateral information to navigate odor sources. We looked at how the model behaves in the presence of one or more odor spot sources, circular and infinite straight trails, and trails with gaps and angles. To allow for mathematical analysis of the model, some simplifications were applied. Instead of using more realistic odor description such as turbulent plumes \cite{Connor2018}, we present concentration as fixed Gaussian distributions. We also keep the function that determines the change in the heading angle linear in the difference between left and right concentration unlike previous work \cite{Calenbuhr:1992ab,Calenbuhr:1992aa}. Calenbuhr et. al \cite{Calenbuhr:1992ab,Calenbuhr:1992aa} put the concentrations through a Michaelis-Menten type nonlinearity so that saturation occurs at large concentrations. These nonlinearities will not change the qualitative behavior (in fact, on an infinite trail, the fixed points are the same), but will alter some of the details like the basin of attraction and the degree of multistability. Some animals change their velocities while searching for odor sources (for example ants \cite{Draft2018} and mice \cite{Liu2019} decrease their velocity closer to the source), here though, we do not take variable velocity into consideration. With our simplifications, we are able to examine how the performance changes as we vary different parameters. The main parameters we look at in our scaled model are the length $l$ of the sensors, the angle $\phi$ between the sensors and the sensitivity $\beta$ to concentration change. 
\\

In the case of the infinite line, as we increase $\beta$, both the analytical and simulated basins of attraction increase which is expected since the change in heading angle becomes more sensitive to the concentration difference. When $\phi$ is larger or $l$ is smaller, we see increased sinusoidal motion centered at the trail.
When the odor source is a spot, one of the fixed points of the model is a saddle point and the other is unstable (at $r$ close to 0). This suggests that the individual will not be able to find spot sources, however, we can see from figures (in $(x,y)$ plane) that trajectories pass through the spot. The reason that the model moves away from the spot is that we have required that the speed stay constant.   
When multiple spot odor sources are added, the $(x,y,\theta)$ system exhibits trajectories that pass through one source or multiple sources, periodic orbits around sources and chaotic behavior.
Because Gaussian circular trails share the radially symmetric property with single spots, we use the same $(r,\xi)$ system to study how varying $\l$, $\beta$ and $\phi$ affects its stability and basin of attraction on these trails.  The fixed point  (circular trail) becomes stable at a small radius ($r_0\sim0.5$) and remains stable for all larger radii. As in infinite trails, when we increase $\beta$ on a circular trail with large enough radius, the basin of attraction increases. This is not true for smaller radii or when we increase the length of the nares $l$ where an optimal length $l<r_0$ gives the largest basin of attraction.  
When the odor source is a finite straight trail, the individual will keep on the trail once it finds it even if there is a gap because of the symmetry between the nares. If the trajectory is sinusoidal (e.g. Fig. \ref{fig:gap}B)  then the individual can lose the trail at the gap depending on the amplitude of the fluctuations. When there is a  trail with an angle, the individual turns and keeps on the trail if the angle is larger than $\pi/4$ and loses the trail otherwise. If the trail bifurcates into two branches, we see that the individual chooses the branch with a smoother turn angle. This is seen in rats \cite{Khan:2012aa} where they tended to choose the branch that had a smaller angle with the main trail (straighter).
\\

At a fixed sensor length $l$, there is an angle $\phi$ that makes the system most stable and have an optimal basin of attraction when the odor landscape is an infinite line. For a spot source, we are also able to find an optimal $\phi$ by freezing the individual while orienting it towards the spot ($\xi =\pi$) and studying the linearization of the new system. We conclude that the individual will best reach the source if it keeps $\phi$ closer to zero when it is away from the spot and closer to $\pi/2$ (large sinusoidal behavior) when it is near the spot. This contradicts the best strategy we found to acquire and stay on an infinite trail where it is better to have a smaller $\phi$ closer to the trail. This shows that animals consider different ways to optimize their search depending on the odor distribution. For example, similar to our results, in Draft et al \cite{Draft2018}, ants move their antennas to have smaller angles while following the trail and bigger angles when exhibiting sinusoidal movements near the trail. In Khan et al \cite{Khan:2012aa}, rats were able to cross gaps and reacquire the trail by increasing the amplitude of their head casting (which might suggest that they are using the strategy discussed to best find infinite trails by changing their method since they can not control angle between nares). Also in Liu et al \cite{Liu2019}, mice exhibit an increase in sinusoidal behavior near the spot source and their trajectories become more tortuous. 
\\

 Real odor landscapes are not simple smooth gradients, but, rather, temporally complicated and turbulent. In Boie et al \cite{JDV,Draft2018}, the authors showed that the spatial information provided by the two sensors is non-redundant in turbulent plumes. We have tested the simple bilateral algorithm in a  plume (not shown here) and we observe that the individual can successfully find the odor source. Similar to our previous results, we have to start at a position orienting towards the plume in order to find it because we do not add noise or a corrective method to turn the individual back once it veers off the plume. One major aspect that we have not explored in this paper is the effects of noise on the models. There are several ways we could introduce this variability in the model. For example, the odor concentration at a point in space could be converted to a rate for a Poisson process and the number of hits in some window of time could act as the main signal. In other work (\cite{Liu2019}), we have used this type of model to mimic the behavior of mice looking for spots of odor.  Another type of stochasticity that could be included is additive noise to the equation for $\theta$. That is, in absence of any odor cue, the individual undergoes a correlated random walk.  Such behavior is commonly seen as a foraging strategy for animals and in the present case would have the effect of allowing the individual to correct for starting conditions that, in the deterministic case, would lead it away from the odor source.  Whether there is an optimal amount of such "noise" to maximize the probability of success is currently a subject of further research. The bilateral model explains many results observed in animal data but not all behavior. Understanding the underlying dynamics of the bilateral model will help in building models that use bilateral information together with other strategies such as casting or upwind orientation.

\bibliographystyle{siamplain}

%\bibliography{finalbib}

\end{document}